\documentclass[%
 preprint,
superscriptaddress,
 amsmath,amssymb,
 aps,
]{revtex4-2}
\usepackage[T1]{fontenc}
\usepackage{graphicx}
\usepackage{dcolumn}
\usepackage{bm}

\begin{document}

\preprint{APS/123-QED}


\title{Manuscript Title:\\Sign reversal of ST-FMR signals in CoFeB/Pt explained by contributions beyond spin-orbit torque}


\author{Christian Berg}
\affiliation{Institute of Physics, Johannes Gutenberg University Mainz, Mainz 55099, Germany}
\affiliation{University of Groningen, 9712 CP Groningen, Netherlands}

\author{Ao Du}
\email{ao.du@univ-lorraine.fr}
\altaffiliation{Present address: Institut Jean Lamour, Université de Lorraine, CNRS, Nancy, 54000, France}
\affiliation{Institute of Physics, Johannes Gutenberg University Mainz, Mainz 55099, Germany}

\author{Durgesh Kumar}
\affiliation{Institute of Physics, Johannes Gutenberg University Mainz, Mainz 55099, Germany}

\author{Sinan Shu}
\affiliation{Institute of Physics, Johannes Gutenberg University Mainz, Mainz 55099, Germany}

\author{Sourabh Manna}
\affiliation{Institute of Physics, Johannes Gutenberg University Mainz, Mainz 55099, Germany}

\author{Sachin Krishnia}
\affiliation{Institute of Physics, Johannes Gutenberg University Mainz, Mainz 55099, Germany}

\author{Mathias Kläui}
\email{klaeui@uni-mainz.de}
\affiliation{Institute of Physics, Johannes Gutenberg University Mainz, Mainz 55099, Germany}
\affiliation{Center for Quantum Spintronics, Norwegian University of Science and Technology, Trondheim 7491, Norway}

\noaffiliation

\date{\today}

\begin{abstract}
Spin-orbit torques (SOT) provide an efficient route for electrical control of magnetization and is of central interest for low-power magnetic memory applications. Spin-torque ferromagnetic resonance (ST-FMR) is a widely used technique to quantify SOT efficiencies. However, the rectified voltage measured in ST-FMR can contain contributions that are not directly related to SOTs, which affects the accuracy of the extracted torque efficiencies. This may contribute to the large variation of reported SOT efficiencies in literature. In this study, we investigate the damping-like torque efficiencies in CoFeB/Pt heterostructures as a function of Pt thickness. Conventional ST-FMR analysis yields an apparent sign reversal of the extracted efficiency at a Pt thickness of approximately 10~nm. By combining complementary measurement geometries and analysis approaches, we separate the torque-induced signal from additional voltage contributions that share the same angular dependence. Our results show that non-torque voltage contributions can produce an apparent sign reversal in conventional ST-FMR analysis, highlighting the need to distinguish SOT-induced signals from parasitic contributions and we show how to reliably determine the SOT efficiencies.
\end{abstract}

\maketitle


\section{\label{sec:INTRODUCTION} INTRODUCTION}

Spintronics exploits the spin degree of freedom of electrons and offers promising routes toward low-power information technologies~\cite{RevModPhys.76.323, doi:10.1126/science.1065389}. A key application is magnetoresistive random access memory (MRAM)~\cite{Chappert_Fert_VanDau_2007, Dieny_Prejbeanu_Garello_Gambardella_Freitas_Lehndorff_Raberg_Ebels_Demokritov_Akerman_et, 9452065, Gupta_Bouard_Kammerbauer_Ledesma-Martin_Bose_Kononenko_Martin__2025}, in which binary information is stored in two distinct magnetic states. Its non-volatility, high speed, high endurance, and low energy consumption make MRAM a promising candidate for next-generation memory technologies. Spin-orbit torques (SOT)~\cite{RevModPhys.91.035004, Miron_Garello_Gaudin_Zermatten_Costache_Auffret_Bandiera_Rodmacq_Schuhl_Gambardella_2011, doi:10.1126/science.1218197, Wang_Cai_Zhu_Wang_Kan_Zhao_Cao_Wang_Zhang_Zhang_et, Du_Zhu_Cao_Zhang_Guo_Shi_Xiong_Xiao_Cai_Yin_et, https://doi.org/10.1002/adma.202505190} and, more recently, orbital torques~\cite{PhysRevLett.125.177201_2020, Go_Jo_Lee_Kläui_Mokrousov_2021, nns5-cn81_2026, Fukami2026, doi:10.1126/science.adw1808} have attracted considerable research interest as a mechanism for MRAM write operations, because they enable energy-efficient electrical control of magnetization. Here, we focus on SOT in conventional heavy metal (HM)/ferromagnet (FM) heterostructures, where an in-plane (IP) charge current flowing through the HM layer generates a transverse spin current via the spin Hall effect (SHE)~\cite{PhysRevLett.83.1834, PhysRevLett.85.393, RevModPhys.87.1213}. The resulting spin current is injected into the adjacent FM layer, where it exerts SOT on the magnetization and can induce switching.

Reliable quantification of SOTs is essential for evaluating the charge-to-spin conversion and optimizing material systems for efficient SOT-based devices. Spin-torque ferromagnetic resonance (ST-FMR) is one of the most widely used techniques for characterizing SOTs in magnetic heterostructures~\cite{Fuchs2007SpintorqueFR, 2011stfmr, 2015stfmrRalph, Skowronski2018DeterminationOS, Shi_Liang_Zhu_Cai_Pollard_Wang_Wang_Wang_He_Yu_et_2019, Chen_Shi_Shi_Fan_Song_Zhou_Bai_Liao_Zhou_Zhang_etal.2021, Lee_Go_Park_Jeong_Ko_Yun_Jo_Lee_Go_Oh_et_2021, Bose2022, PhysRevB.107.134423_2023}. In ST-FMR, current-induced torques drive the magnetization into resonance, and the resulting rectified voltage is analyzed to determine the effective torques acting on the magnetic layer. However, the measured voltage can also contain contributions that do not directly reflect the SOT efficiency of the HM layer, such as the self-torque of the FM layer, spin pumping and thermoelectric effects~\cite{PhysRevB.104.094430, 10.1063/5.0038567, Bose2018SpinNernst, Liu_Zhu_2025}. These contributions may appear in the same symmetric Lorentzian component of the ST-FMR spectra and may exhibit the same dependence on magnetic field angle as the torque-induced signal. Consequently, conventional lineshape analysis may yield inaccurate values of the extracted SOT efficiency. While Pt is among the most extensively studied heavy metals for SOT generation, reported SOT efficiencies span a broad range~\cite{Liu2011ReviewAA, Wang_Deorani_Qiu_Kwon_Yang_2014, 2015stfmrRalph,  Zhang_Han_Jiang_Yang_Parkin_2015, lineshapevslinewidth}. This variation highlights the need to carefully assess the reliability of extracted SOT efficiencies using conventional ST-FMR measurements. 

In this study, we use ST-FMR to systematically investigate the Pt thickness dependence of the damping-like torque (DLT) efficiency in CoFeB/Pt heterostructures. Conventional lineshape analysis yields an efficiency that decreases monotonically with increasing Pt thickness and exhibits a sign reversal at approximately 10~nm. To elucidate the origin of this behavior, we perform both longitudinal and transverse ST-FMR measurements to separate the torque-induced response from additional voltage contributions that exhibit the same dependence on the applied-field angle. After applying a correction method, the extracted torque efficiency remains positive for all Pt thicknesses investigated, indicating that the apparent sign reversal does not originate from an intrinsic reversal of the SOT sign. This conclusion is further supported by direct-current (DC) modulated damping measurements~\cite{2011stfmr}. Our findings show that parasitic voltages can mimic a reversal of the SOT and highlight the importance of complementary measurement geometries for the reliable determination of charge-to-spin conversion.

\section{\label{sec:METHODS} METHODS}

Samples were deposited by magnetron sputtering on thermally oxidized Si substrates with 300~nm SiO$_\text{x}$ using a Singulus Rotaris deposition system. The sample stacks were Sub.//$\text{Co}_{20}\text{Fe}_{60}\text{B}_{20}$($t_{\text{CoFeB}}$)/Pt($t_{\text{Pt}}$) and Sub.//$\text{Ni}_{81}\text{Fe}_{19}$($t_{\text{Py}}$)/Pt($t_{\text{Pt}}$), where $\text{Co}_{20}\text{Fe}_{60}\text{B}_{20}$ and $\text{Ni}_{81}\text{Fe}_{19}$ are hereafter denoted as CoFeB and Py, respectively, and numbers in parentheses denote nominal layer thicknesses in nanometers. The deposited films were patterned into devices using a two-step optical lithography process. We employed three complementary ST-FMR signal analysis approaches on these devices: conventional lineshape analysis, combined longitudinal and transverse measurements, and DC-induced linewidth modulation. The corresponding measurement geometries and results are presented in Figures~\ref{fig:figure1}, \ref{fig:figure2} and~\ref{fig:figure4}.

\subsection{\label{sec:conventional} Conventional Lineshape Analysis}
The damping-like SOT efficiency extracted using conventional lineshape analysis is denoted $\xi_{\mathrm{FMR}}$. In this analysis, the measured rectified mixing voltage $V_{\mathrm{mix}}$ results from a homodyne mixture of the radio-frequency (RF) current with resistance oscillations arising from anisotropic magnetoresistance (AMR) and is given by~\cite{2011stfmr}

\begin{equation}
    V_{\mathrm{mix}} \propto S(\phi) \frac{\Delta^2}{(H - H_{\mathrm{res}})^2 + \Delta^2} + A(\phi) \frac{\Delta( H - H_{\mathrm{res}} )}{(H - H_{\mathrm{res}})^2 + \Delta^2},
    \label{stfmrlineshape}
\end{equation}

where $S$ and $A$ are the amplitudes of the symmetric and antisymmetric Lorentzians, respectively, $H$ is the applied magnetic field, $H_{\mathrm{res}}$ is the resonance field, and $\Delta$ is the half-width-at-half-maximum. The IP field angle was set to $\phi = 45^\circ$ relative to the RF current direction, which maximizes the AMR sensitivity. When the field-like torque makes only a small contribution to the antisymmetric Lorentzian~\cite{Skowronski2018DeterminationOS}, the FMR efficiency $\xi_{\mathrm{FMR}}$ can be calculated as  

\begin{equation}
    \xi_{\mathrm{FMR}} = \frac{S}{A} \frac{e \mu_0 M_{\mathrm{s}} t_{\mathrm{FM}} t_{\mathrm{HM}}}{\hbar} \sqrt{1 + \frac{\mu_0 M_{\mathrm{eff}}}{H_{\mathrm{res}}}}, 
    \label{sotefficiency}
\end{equation}

where $e$ is the elementary charge, $\mu_0$ is the permeability of free space, $M_{\mathrm{s}}$ is the saturation magnetization, $t_{\mathrm{FM}}$ and $t_{\mathrm{HM}}$ are the thicknesses of the FM and HM layers, respectively, $\hbar$ is the reduced Planck's constant, and $\mu_0 M_{\mathrm{eff}}$ is the effective magnetization~\cite{2011stfmr}. 

\subsection{\label{sec:non-torque} Separation of Torque and Non-Torque Contributions}

In-plane angular measurements were performed by recording ST-FMR spectra as a function of the angle $\phi$ between the equilibrium magnetization and the RF current. In our coordinate system, the RF current flows along the $x$ axis, the $y$ axis lies in the film plane perpendicular to the current, and the $z$ axis is normal to the film plane. In the longitudinal geometry, the rectified mixing voltage is detected along the current path (yielding signals denoted with the subscript $XX$. Here, the symmetric Lorentzian amplitude $S_{xx}$ follows a $\sin(2\phi)\cos\phi$ dependence~\cite{AngularScans}. However, voltage contributions from spin pumping~\cite{PhysRevB.82.214403} and thermoelectric effects~\cite{PhysRevB.90.224427} can exhibit the same angular dependence. Therefore, longitudinal ST-FMR measurements alone cannot unambiguously distinguish the torque-induced signal from these additional voltage contributions. 

To resolve this ambiguity, we combine longitudinal with transverse ST-FMR measurements, where the rectified voltage is detected across Hall probes along the $y$ axis (denoted with the subscript $XY$. Measuring both geometries provides the necessary symmetry information required for the separation of torque-driven and parasitic voltage signals. By incorporating the non-torque contributions in the fitting coefficients, the angular dependencies of the symmetric Lorentzian $S_{XX}$ and antisymmetric Lorentzian $A_{XX}$ are expressed as~\cite{Free_energy_to_Heff_2020DR, ZrTe3paper}

\begin{equation}
    S_{XX}(\phi) = S_{XX}^{\text{AMR/nt}} \sin 2\phi \cos \phi + S_{XX}^{\text{FLT, AMR}} \sin 2\phi ,
    \label{angular1}
\end{equation}
\begin{equation}
    A_{XX}(\phi) = A_{XX}^{\text{AMR}} \sin 2\phi \cos \phi + A_{XX}^{\text{DLT, AMR}} \sin 2\phi.
    \label{angular2}
\end{equation}

The transverse rectified voltage arises from the mixing of the RF current with the resistance oscillations associated with the planar Hall effect (PHE) and anomalous Hall effect (AHE) resistances. Its symmetric and antisymmetric amplitudes are written as

\begin{equation}
    S_{XY}(\phi) = S_{XY}^{\text{PHE/nt}} \cos 2\phi \cos \phi + S_{XY}^{\text{AHE/nt}} \cos \phi + S_{XY}^{\text{FLT, PHE}} \cos 2\phi + S_{XY}^{\text{DLT, AHE}},
    \label{angular3}
\end{equation}
\begin{equation}
    A_{XY}(\phi) = A_{XY}^{\text{PHE}} \cos 2\phi \cos \phi + A_{XY}^{\text{AHE}} \cos \phi + A_{XY}^{\text{DLT, PHE}} \cos 2\phi + A_{XY}^{\text{FLT, AHE}}.
    \label{angular4}
\end{equation}

In these equations, the superscript \textit{nt} identifies coefficients containing non-torque voltage contributions.


The coefficient $A_{XY}^{\text{AHE}}$ is free from non-torque contributions and thus depends primarily on the IP DLT. The coefficients $S_{XX}^{\text{AMR/nt}}$ and $S_{XY}^{\text{PHE/nt}}$ contain both this torque-driven component and parasitic non-torque contributions. All fitting parameters $S$ and $A$ associated with the AMR, PHE, and AHE can be used to calculate the non-torque voltage $V_{nt}$ from the following two equations: 

\begin{equation}
    \eta = \frac{-A_{XY}^{\text{AHE}}}{S_{XY}^{\text{AHE/nt}} + V_{\text{nt}}} = 
    \begin{cases} 
    \dfrac{S_{XY}^{\text{PHE/nt}} + V_{\text{nt}}}{A_{XY}^{\text{PHE}}}, \\[15pt]
    \dfrac{S_{XX}^{\text{AMR/nt}} + V_{\text{nt}}}{A_{XX}^{\text{AMR}}}.
    \end{cases}
    \label{eta}
\end{equation}

Once the non-torque voltage contribution is determined, the corrected FMR efficiency can be extracted. In this procedure, the ratio $\frac{S}{A}$ in Equation~\ref{sotefficiency} is replaced with $\eta$ in Equation~\ref{eta}. Because $\eta$ can be obtained independently from both the AMR-based and PHE-based relations in Equation~\ref{eta}, comparing the efficiencies derived from these two expressions serves as a consistency check for the correction procedure.

\subsection{\label{sec:DCmodulated} DC-Induced Linewidth Modulation}

Alternatively, the DLT efficiency can be determined from the DC-current-induced modulation of the resonance linewidth. A DC current is applied to the device in addition to the RF excitation. The spin current generated through the SHE exerts a DLT on the adjacent FM layer, modifying its effective damping. The resonance linewidth and effective damping parameter vary linearly with the charge-current density $J_{C}$~\cite{10.1063/1.4999948}. The damping parameter is given by

\begin{equation}
    \alpha = \alpha_0 + \Delta \alpha(J_{\mathrm{c}}) = \alpha_0 + \frac{\sin(\phi)}{\mu_0 M_{\mathrm{s}} t_{\mathrm{FM}}\left(H_{\mathrm{ext}} + \frac{M_{\mathrm{eff}}}{2}\right)} \frac{\hbar}{2e} J_{\mathrm{c}} \xi_{\mathrm{DLT}},
    \label{dcmodulated}
\end{equation}

where $\alpha_0$ is the effective damping parameter at zero DC current, and $\Delta \alpha(J_c)$ is the current-induced change in damping~\cite{2011stfmr, PhysRevB.91.214416, lineshapevslinewidth}. The DLT efficiency is extracted from the slope of $\alpha$ versus $J_c$. Reversing the applied magnetic field direction reverses the relative orientation between the magnetization and spin polarization and is therefore expected to reverse the sign of the slope. This method determines the torque efficiency from a current-induced change in linewidth, rather than from the ratio of symmetric and antisymmetric voltage amplitudes.

\section{\label{sec:RESULTS} RESULTS AND DISCUSSION}

To assess the robustness of the analysis methods, we compare the results of the three complementary measurement and analysis approaches. First, we determine the FMR efficiency as a function of Pt thickness in CoFeB/Pt and Py/Pt heterostructures using conventional ST-FMR lineshape analysis. We then investigate the origin of the observed behavior through angular longitudinal and transverse measurements. Finally, we validate the interpretation using DC-modulated linewidth measurements.

\subsection{\label{sec:ConventionalResults} Apparent Sign Reversal in Conventional Lineshape Analysis}

\begin{figure}[hbtp]
    \centering
    \includegraphics[width=1\linewidth]{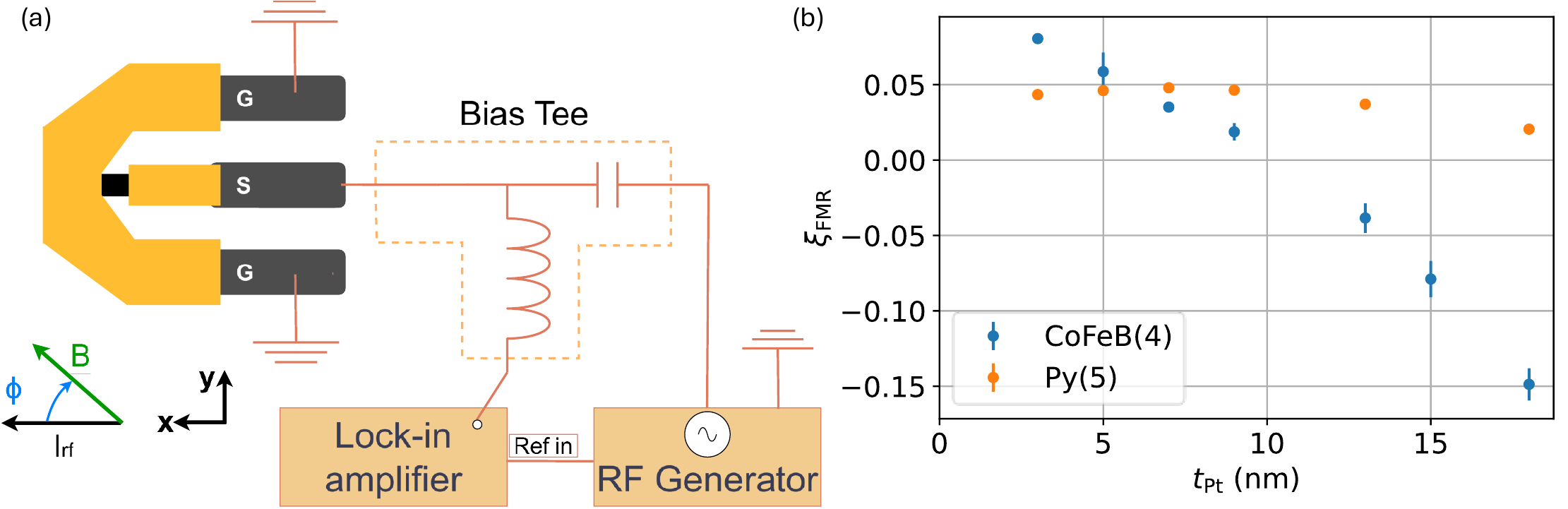}
    \caption{(a) Schematic of the conventional longitudinal ST-FMR measurement geometry. The measured rectified DC voltage originates from a homodyne mixture of the RF current with the oscillating AMR. (b) FMR efficiency extracted using the conventional ST-FMR lineshape analysis for the CoFeB/Pt (in blue) and Py/Pt (in orange) series as a function of Pt thickness. A sign reversal near 10 nm is observed only in the CoFeB/Pt series.}
    \label{fig:figure1}
\end{figure}

We investigated two sample series, CoFeB(4)/Pt($t_{\text{Pt}}$) and Py(5)/Pt($t_{\text{Pt}}$), with Pt thicknesses ranging from 3 to 18~nm. The ST-FMR spectra were analyzed with the conventional lineshape analysis in the measurement geometry shown in Figure~\ref{fig:figure1}(a). Within each series, the extracted effective magnetization $M_{\mathrm{eff}}$ and damping parameter $\alpha$ remain constant with Pt thickness, see SI1. Figure~\ref{fig:figure1}(b) shows the FMR efficiencies versus Pt thickness. 

Because the spin diffusion length in Pt is about 1.5~nm~\cite{PhysRevB.91.024402}, the DLT efficiency generated by the SHE is expected to approach saturation when the Pt thickness is a few times the spin diffusion length. Contrary to this expectation, the apparent FMR efficiency extracted from the CoFeB/Pt series decreases with increasing Pt thickness and reverses sign at approximately 10~nm. Representative lineshapes, exhibiting opposite signs for the symmetric components, are shown in SI2. The FMR efficiency remains positive throughout the same Pt-thickness range in the Py/Pt reference series, as shown in Figure~\ref{fig:figure1}(b). Because the SHE in Pt generates a spin current of the same sign in both series, the contrast between CoFeB/Pt and Py/Pt suggests that the sign reversal is not due to the Pt-generated spin current. Instead, it may arise from additional voltage contributions associated with the FM layer or interface, whose relative importance depends on the magnitude of the magnetoresistive mixing signal. 

\subsection{\label{sec:non-torquecorrectionResults} Correction for Contributions Beyond Spin-Orbit Torque}
\begin{figure}[hbtp]
    \centering
    \includegraphics[width=1\linewidth]{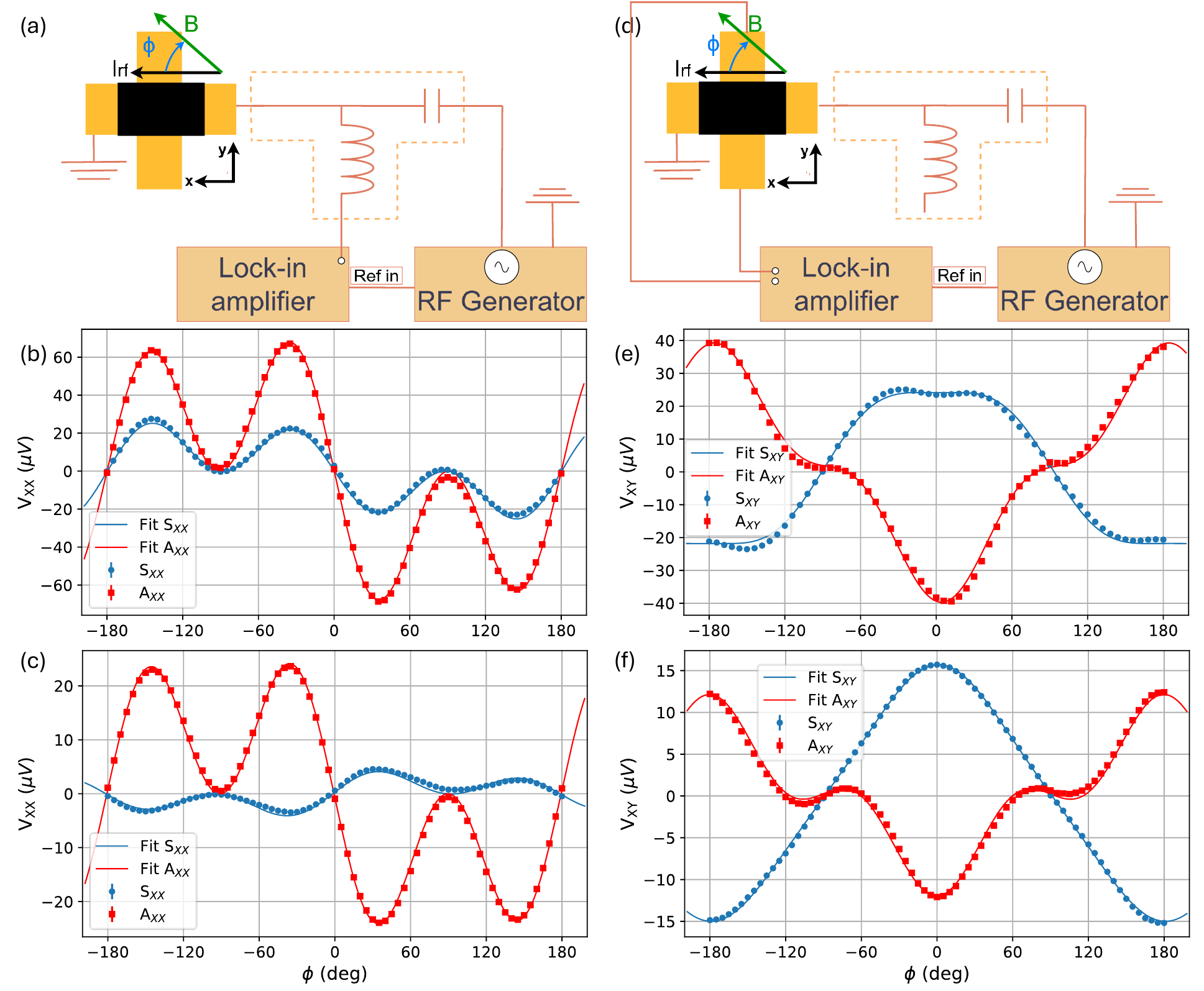}
    \caption{(a) Longitudinal ST-FMR measurement geometry. Angular dependencies of $S_{XX}$ and $A_{XX}$ for (b) CoFeB(4)/Pt(5) and (c) CoFeB(4)/Pt(15). (d) Transverse ST-FMR measurement geometry. Angular dependencies of $S_{XY}$ and $A_{XY}$ for (e) CoFeB(4)/Pt(5) and (f) CoFeB(4)/Pt(15).}
    \label{fig:figure2}
\end{figure}

To identify the voltage contributions responsible for the apparent sign reversal, we performed angular dependent measurements in longitudinal and transverse geometries. We first performed longitudinal measurements on the CoFeB sample series, using the measurement geometry shown in Figure~\ref{fig:figure2}(a). Two angular scans for Pt thicknesses of 5 and 15~nm are shown in Figures~\ref{fig:figure2}(b) and~\ref{fig:figure2}(c), respectively, together with fits using Equations~\ref{angular1} and~\ref{angular2}. For CoFeB(4)/Pt(5), the symmetric Lorentzian component $S$ follows the expected $\sin(2\phi)\cos(\phi)$ dependence. The CoFeB(4)/Pt(15) sample exhibits the same angular dependence, but the symmetric Lorentzian component $S$ has opposite sign. Since $\xi_{\mathrm{FMR}} \propto S/A$ in the conventional lineshape analysis, the sign reversal yields an apparent negative FMR efficiency. 

The absence of a sign reversal in the Py/Pt series indicates that the apparent sign reversal observed in CoFeB/Pt is unlikely to originate from a reversal of the spin current polarity generated in Pt. Instead, the observed behavior may be associated with additional voltage contributions that depend on the FM layer or interface. However, because such contributions can exhibit the same angular dependence as the torque-induced signal, longitudinal ST-FMR measurements alone cannot unambiguously determine the origin of the apparent sign reversal in CoFeB/Pt. Measurements of a CoFeB thickness series indicate that self-torques in CoFeB do not contribute substantially to the measured voltage, see SI3. We therefore performed transverse ST-FMR measurements using the geometry shown in Figure~\ref{fig:figure2}(d). The resulting PHE- and AHE-related coefficients, combined with the longitudinal AMR-related coefficients, allow the torque-induced response to be separated from non-torque voltage contributions using Equations~\ref{eta}. The transverse angular scans for Pt thicknesses of 5 and 15~nm are shown in Figures~\ref{fig:figure2}(e) and~\ref{fig:figure2}(f), respectively, together with the fits based on Equations~\ref{angular3} and~\ref{angular4}. 

\begin{figure}[hbtp]
    \centering
    \includegraphics[width=1\linewidth]{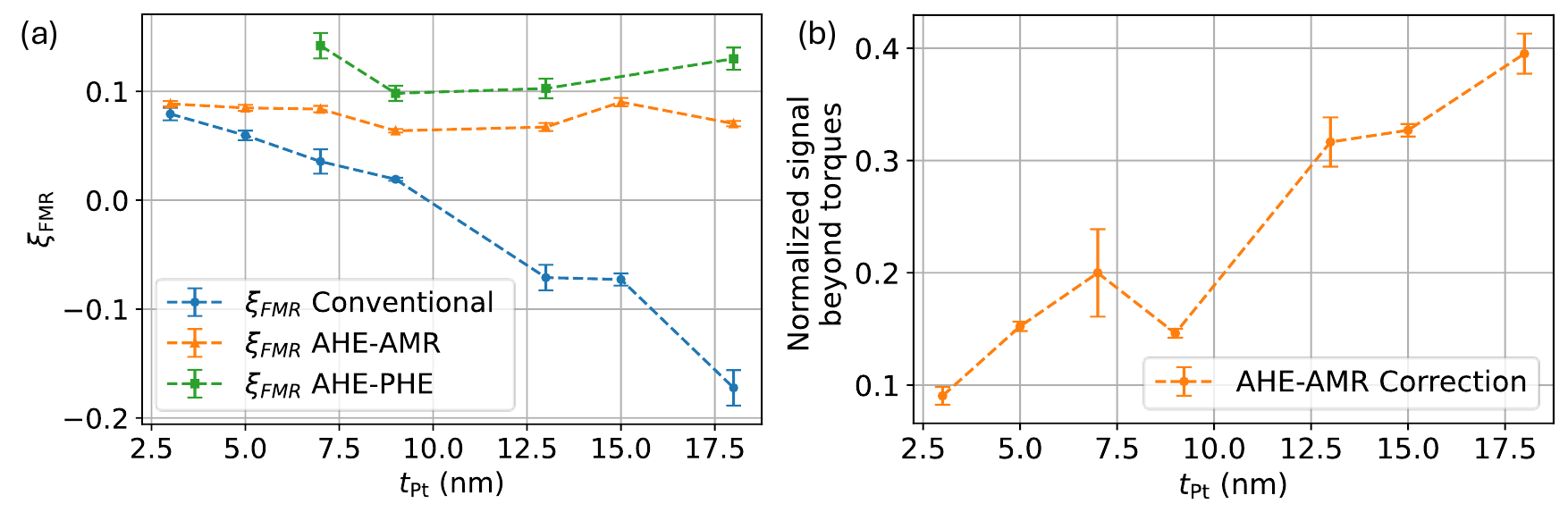}
    \caption{(a) The calculated FMR efficiency using different methods: conventional lineshape analysis (blue), AHE-AMR correction (orange) and AHE-PHE correction (green). (b) Non-torque voltage contribution based on the AHE-AMR correction equation, normalized by $A_{XX}^{AMR}$. }
    \label{fig:figure3}
\end{figure}

Figure~\ref{fig:figure3}(a) compares the corrected FMR efficiencies with the values obtained from conventional lineshape analysis, which contains contributions beyond SOT. The uncorrected efficiency decreases with increasing Pt thickness and becomes negative for the thicker Pt layers (blue in Figure \ref{fig:figure3}(a)). In contrast, after correction for non-torque voltage contributions, the extracted efficiencies remain positive. Both the AHE-AMR (orange in Figure \ref{fig:figure3}(a)) and AHE-PHE (green in Figure \ref{fig:figure3}(a)) correction methods yield similar efficiencies, supporting the consistency of the extraction procedure~\cite{Free_energy_to_Heff_2020DR, ZrTe3paper}. This result indicates that the apparent sign reversal observed in the conventional analysis does not reflect an intrinsic reversal of the DLT. These non-torque voltages may arise from spin pumping and thermoelectric effects, because their contributions can exhibit the same angular dependence as the torque-induced rectification signal~\cite{Free_energy_to_Heff_2020DR, ZrTe3paper}. 

The non-torque voltage contributions calculated from the AMR response are normalized by the denominator $A^{AMR}_{XX}$ from Equation~\ref{eta} and then they are plotted in Figure~\ref{fig:figure3}(b). The normalized contribution generally increases with Pt thickness. This trend supports the interpretation that additional voltage contributions increasingly affect the conventional ST-FMR signal in samples with thicker Pt layers, eventually producing an apparent sign reversal.

The contrasting behavior between the CoFeB/Pt and Py/Pt series in Figure~\ref{fig:figure1}(b) is also consistent with their different AMR signal responses (SI4). The larger AMR ratio in Py/Pt produces a stronger torque-induced rectified voltage, thereby reducing the relative influence of non-torque voltage contributions that are not proportional to the AMR. This provides a possible explanation for the positive apparent efficiency retained in Py/Pt throughout the investigated thickness range. In contrast, the weaker AMR response in CoFeB/Pt, especially at larger Pt thicknesses where current shunting reduces the torque-induced mixing voltage, allows the non-torque contributions to dominate and eventually produce an apparent sign reversal.

\subsection{\label{sec:DCmodulatedResults} Torque Efficiencies from DC-Induced Linewidth Modulation}

\begin{figure}[hbtp]
    \centering
    \includegraphics[width=1\linewidth]{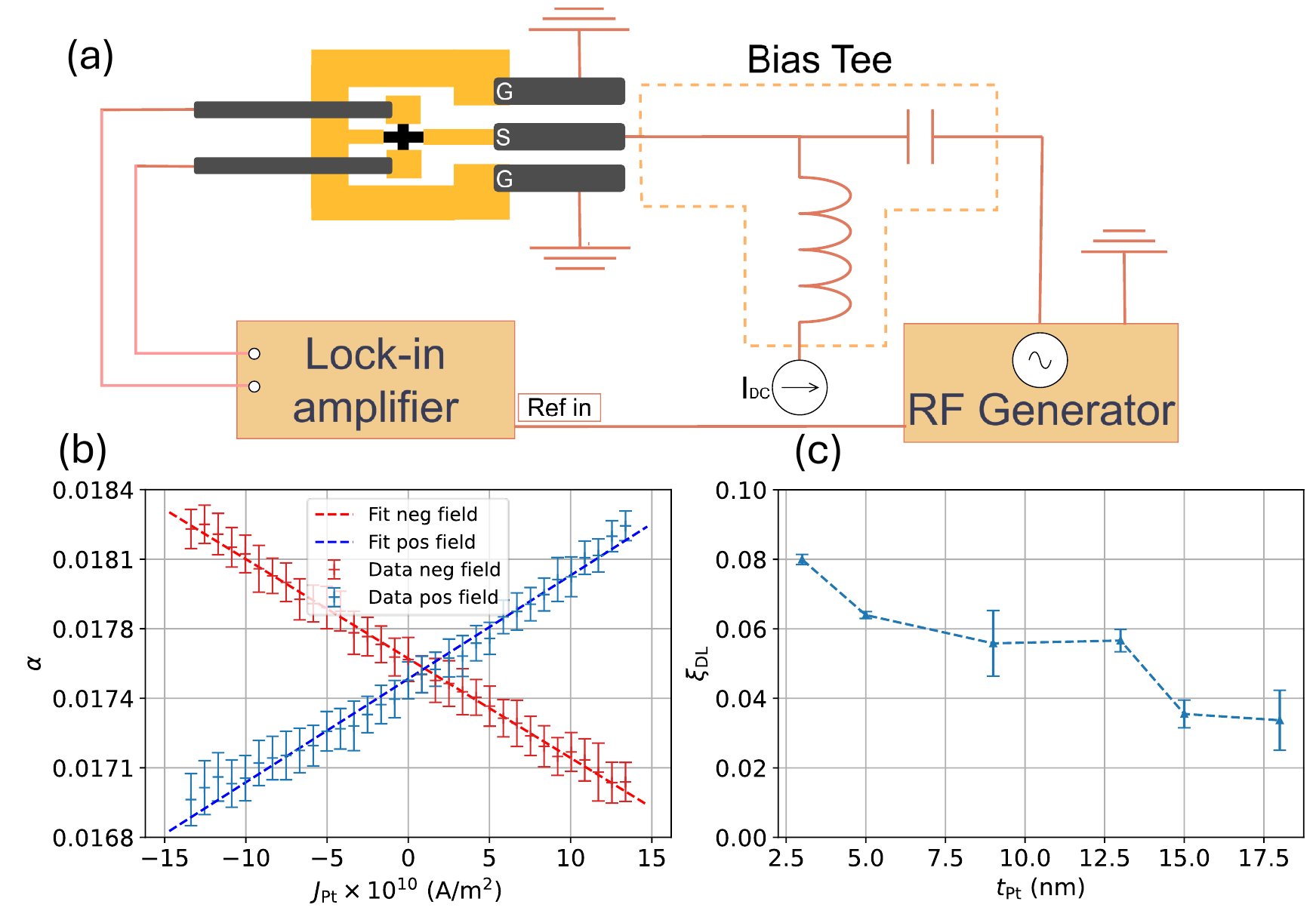}
    \caption{(a) Schematic of the measurement setup used for DC-induced linewidth modulation measurements. (b) Dependence of the damping parameter on the applied direct current density for the two opposite field directions for CoFeB(4)/Pt(5). (c) Extracted DLT efficiency versus Pt thickness for CoFeB(4)/Pt($t_{\text{Pt}}$), the DLT efficiency is positive for all Pt thicknesses.}
    \label{fig:figure4}
\end{figure}

To obtain an independent determination of the DLT sign, we performed DC-induced linewidth modulation measurements using the schematic diagram shown in Figure~\ref{fig:figure4}(a). For CoFeB(4)/Pt(5) in Figure~\ref{fig:figure4}(b), the extracted damping parameter decreases linearly with current density for one field direction and increases with a slope of equal magnitude but opposite sign for the opposite field direction. 

The torque efficiencies extracted using Equation~\ref{dcmodulated}, averaged over the two field directions, are shown in Figure~\ref{fig:figure4}(c). Although the extracted efficiencies decrease slightly with increasing Pt thickness, they remain positive over the full thickness range investigated. In particular, samples for which the conventional lineshape analysis yields negative apparent efficiencies retain a positive efficiency in the DC-modulated measurements. The agreement in torque sign between the DC-modulated measurements and the corrected voltage analysis corroborates the conclusion that the conventional ST-FMR sign reversal arises from additional voltage contributions rather than an intrinsic reversal of the DLT sign. 

\section{\label{sec:CONCLUSION} CONCLUSION}

We investigated and compared the Pt thickness dependence of the damping-like SOT efficiency in CoFeB/Pt and Py/Pt heterostructures using conventional ST-FMR lineshape analysis, longitudinal and transverse ST-FMR measurements, and DC-induced linewidth modulation.

An apparent sign reversal of the SOT efficiency appears only in thicker CoFeB/Pt samples analyzed via conventional lineshape analysis. The sign reversal is absent in Py-based samples, which implies that the apparent signal reversal does not arise from an intrinsic reversal of SOT sign. The analysis indicates that non-torque contributions, such as spin pumping and thermally induced voltages, share the angular dependence of the symmetric Lorentzian component used to extract the FMR efficiency in the conventional analysis. As the Pt thickness increases, the mixing voltage of the structure decreases significantly owing to current shunting effects, thereby reducing the mixing voltage generated by SOT. Consequently, other voltage contributions become dominant, leading to an inaccurate determination of the FMR efficiency when using the conventional lineshape method.

Our findings highlight the limitation of conventional ST-FMR lineshape analysis in the thicker Pt structures studied here, particularly in samples with comparatively weak AMR. The results demonstrate that additional measurement techniques and ST-FMR corrections are essential for a reliable determination of SOT efficiencies. 

\newpage

\begin{acknowledgments}                                                                                                                                                                                                                                                                                                                                                                 
The authors acknowledge fruitful discussions with K. Ando from Keio University. The authors acknowledge funding by the DFG (TRR 173-268565370 Spin+X: A01, A11, B02; TRR 288-422213477 Elasto-Q-Mat: A12 and project 358671374), the Horizon Europe Framework (grant number 101070290 (NIMFEIA), grant number 101070287 (Swan-on-Chip), grant number 101226840 (ORBIS) and grant number 101129641 (OBELIX)), the European Research Council (grant number 856538 (3D MAGiC)), King Abdullah University of Science and Technology (KAUST) under award 2024-CRG12-6480 and the Research Council of Norway through its Centers of Excellence funding scheme under project number 262633 'QuSpin'. A. Du acknowledges the Alexander von Humboldt Foundation Research Fellowship for Postdoctoral Researchers (No. 1261416).


\end{acknowledgments}

\clearpage
\makeatletter
\def\doibase{}
\def\da@href#1#2{#2}
\makeatother
\bibliography{apssamp}

\end{document}